\documentclass{article}
\usepackage{amssymb}
\usepackage{caption}
\usepackage{amsfonts}
\usepackage{amsmath}
\usepackage{fancyhdr}
\usepackage{lastpage}
\usepackage{bbm}
\usepackage{booktabs}
\usepackage{url}
\usepackage{graphicx}
\usepackage{hyperref}
\usepackage[paperheight=11in,paperwidth=8.5in,portrait,left=2.5cm,right=2.5cm,top=2.5cm,bottom=2cm]{geometry}
\usepackage{lipsum}
\usepackage{empheq}
\usepackage{tabularx}
\title{Predicting Earth's Carrying capacity of Human Population as the Predator and the Natural Resources as the Prey in the Modified Lotka-Volterra Equations with Time-dependent Parameters}
\author{Cheng Sok Kin, Ian Man Ut, Lo Hang, U Ieng Hou, Ng Ka Weng, Un Soi Ha, Lei Ka Hin
	\and 
	Cheng Kun Heng, Tam Seak Tim, Chan Iong Kuai,
	Lee Wei Shan\footnote{email: \href{mailto:WSLEEemails}{weishan\_lee@yahoo.com}}
}
\date{Escola Choi Nong Chi Tai\\ Macao Special Administrative Region, People's Republic of China.} 
\begin{document}
	\maketitle
	
	\abstract{We modified the Lotka-Volterra Equations with the assumption that two of the original four constant parameters in the traditional equations are time-dependent. In the first place, we assumed that the human population (borrowed from the T-Function) plays the role as the prey while all lethal factors that jeopardize the existence of the human race as the predator. Although we could still calculate the time-dependent lethal function, the idea of treating the lethal factors as the prey was too general to recognize the meaning of them. Hence, in the second part of the modified Lotka-Volterra Equations, we exchanged the roles between the prey and the predator. This time, we treated the prey as the natural resources while the predator as the human population (still borrowed from the T-Function). After carefully choosing appropriate parameters to match the maximum carrying capacity with the saturated number of the human population predicted by the T-Function, we successfully calculated the natural resources as a function of time. Contrary to our intuition, the carrying capacity is constant over time rather than a time-varying function, with the constant value of 10.2 billion people.}
	\newtheorem{theorem}{Theorem}[section]
	\newtheorem{definition}{Definition}[section]
	\newtheorem{lemma}{definition}[section]
	\pagenumbering{gobble}
	\thispagestyle{empty}
	
	\pagenumbering{arabic}
	\pagestyle{fancy}
	\fancyhf{}  \rhead{Page \thepage\ of \pageref{LastPage}}
	\lhead{Cheng Sok Kin et. al. Predicting Earth's Carrying capacity of Human Population$\cdots$}
	\cfoot{\thepage}
	\renewcommand{\headrulewidth}{1pt}
	
	\section{Introduction}
	From the origin of humankind, there has been such a vast amount of human beings in the world right now. The emerge of globalization, and the development of society have set off magnificent economic progress, generating materials and luxury of human beings' life. Meanwhile, the progress and its associated advantages have forced an enormous cost to the circumstance that we live.\\
	Carrying capacity is the maximum number of species that allows us to stay in the earth$^{\cite{Def 1}}$, and it depends on the conditions and resources available in the specific area, as well as the consumption habits of the species considered$^{\cite{Def 2}}$. Carrying capacity is a measure of sustainability within these changing conditions. It is a conventional believe that because both resources and consumption that are available in the area change over time, carrying capacity is always changing. In addition, Joel et al.$^{\cite{Cohen BESA}}$, as well as Pulliam et al.$^{\cite{Pulliam et. al.}}$ introduced the concept about carrying capacity of human populations in the 1960s. It was noted that the consumption habits of humans are much more variable than those of other animal species, making it considerably more difficult to predict the carrying capacity for human beings. This realization gave rise to the IPAT Equation which pointed out that carrying capacity for humans was a function not only of population size but also of differing levels of consumption, which in turn are affected by the technologies involved in production and utilization. There have been a large number of published estimates for the human carrying capacity of the planet$^{\cite{Cohen Science}-\cite{Hopfenberg}}$; they range from a low of one half billion people to a staggering 800 billion. Many of these estimates are more ideologically based than determined by scientific principles. These exercises demonstrate the complexity of developing useful estimates of the human carrying capacity of the earth, and the limitations of using the methodology which has been successful with non-human species. Also claimed by Cohen$^{\cite{Cohen Science}}$, the carrying capacity of this planet is determined by different factors, namely the restraints of natural circumstance and society element-economic. Moreover, Mohammad$^{\cite{Mohammad}}$ pointed out that carrying capacity is a mathematical concept that assumes the limit of the size of the population and that the carrying capacity of a resource system mostly relies on the size of the needs of that population. The size of the obligation cannot exceed the limit of carrying capacity to maintain sustainability. Furthermore, critical factors for manipulating needs are population numbers, density, affluence, technology, depletion rate of renewable and nonrenewable resources, and finally the build-up of hazardous wastes in the environment. What is more, Hopfenberg et al.$^{\cite{Hopfenberg}}$ considered that the carrying capacity is controlled by food availability and argued that data from Cohen may not be accurate. Furthermore, the Lotka-Volterra equations$^{\cite{Evans and Findley}}$ were also used to describe the relationships between the preys and the predators, making it possible to model the relationships between humans and natural resources. In addition, Taagepera claimed that the "tamed quasi-hyperbolic function", or simply put, the T-Function$^{\cite{T-Function}}$, may successfully describe the human population over time in the range between the 5th century to the 20th century, also predicting the maximum value of human population that will reach the saturation around year 2080.\\
	Nevertheless, none of the research discussed on the calculation of the carrying capacity about humans that are able to please our mind. In addition, the factors they concerned are not enough as a result of plenty unwanted elements that can affect the data of the carrying capacity.\\
	In this study, we aimed to use modified the Lotka-Volterra Equations with the assumption that two of the original four parameters in the traditional equations are time dependent. In the first place, we assumed that the human population plays the role as the prey while all lethal factors that jeopardize the existence of the human race as the predator. Second, we exchanged the roles between the prey and the predator, treating the prey as the natural resources while the predator as the human population. In both cases, we calculate the correspoding time-varying parameters and explain the respective meaning. Contray to our intuition, the carrying capacity is a constant function (with the value $10.2$ billion) rather than a function that changes with time.	
	\section{Theorems and Models}
	We first review the traditional Lotka-Volterra equations with all the parameters all being constants. Second, because the T-Function fits well the human population vs. time, we conjecture that one of the solutions to the equations should satisfy the T-Function. But if we still kept the original parameters constant, the T-Function would not be the solution. Therefore, after reviewing the traditional Lotka-Volterra equations, we ease the restriction on the time-independent parameters to allow two of them to be time-dependent, while not making the equations too difficult to solve. 
	\subsection{Review on Lotka-Volterra equations. }\label{L-V Section}
	The "traditional" Lotka-Volterra equations, also known as the prey and the predator equations, are successfully used to describe the interactions between the predators and preys in the natural system. In the first place, we follow closely the derivations in Ref$\cite{Evans and Findley}$ and Ref$\cite{L-V wiki}$ and review some imporant properties on these equations. The model of these equations demonstrates the trend or the variety of the quantities of both of the preys and the predators to help compare them easily. Further, it can also learn how the population of each side goes when there is the participation of the opposite group. The first equation describes the conditions of the preys, denoted by $x_{1}$, while the second equation describes the conditions of the predator, denoted by $x_{2}$, with the following differential equations:
	\begin{subequations}\label{traditional L-V}
		\begin{empheq}[left=\empheqlbrace]{align}
			\frac{dx_{1}}{dt}&=\alpha x_{1} - \beta x_{1}x_{2};\label{diff for prey}\\	
			\frac{dx_{2}}{dt}&=-\gamma x_{2} + \delta x_{1}x_{2},\label{diff for predator}
		\end{empheq}
	\end{subequations}
	where $\alpha$, $\beta$, $\gamma$, and $\delta$ are all positive constants. 
	From Eq.($1$), we get 
	\begin{center}
		\begin{equation}
		\frac{dx_{1}}{dx_{2}}=\frac{\alpha x_{1}-\beta x_{1}x_{2}}{-\gamma x_{2}+\delta x_{1}x_{2}}
		=-\frac{x_{1}}{x_{2}}\cdot\frac{\alpha-\beta x_{2}}{\gamma - \delta x_{1}}.	
		\end{equation}
	\end{center}
	Standard calculations on the separation of variables lead to 
	\begin{center}
		\begin{equation}\label{diff_form}
		\frac{\gamma-\delta x_{1}}{x_{1}}dx_{1}+\frac{\alpha-\beta x_{2}}{x_{2}}dx_{2}=0.
		\end{equation}
	\end{center}
	Integrating the equation, we get
	\begin{equation}\label{traditional Lambda}
	\gamma\ln(x_{1})-\delta x_{1} + \alpha\ln(x_{2})-\beta x_{2} = -\Lambda = \mathbf{constant}.
	\end{equation}
	Define the Carrying capacity $K$ as
	\begin{equation}\label{carrying capacity}
	K=e^{-\Lambda}=x_{2}^{\alpha}e^{-\beta x_{2}}x_{1}^{\gamma}e^{-\delta x_{1}},
	\end{equation}
	with the maximum value $K^{*}$ at the stationary point $\Bigl(\gamma/\delta,\alpha/\beta\Bigr)$ to be 
	\begin{equation}\label{maximum carrying capacity}
	K^{*}=\Bigl(\frac{\alpha}{\beta e}\Bigr)^{\alpha}\Bigl(\frac{\gamma}{\delta e}\Bigr)^{\gamma}.
	\end{equation}
	Further, if we introduce new coordinates $z_{1}$ and $z_{2}$ such that 
	\begin{subequations}\label{coord trans}
		\begin{empheq}[left=\empheqlbrace]{align}
			z_{1}&=\frac{1}{\sqrt{\Lambda}}(\delta x_{1} + \beta x_{2})^{\frac{1}{2}};\label{z1t}\\	
			z_{2}&=\frac{1}{\sqrt{\Lambda}}(-\gamma\ln{x_{1}} - \alpha\ln{x_{2}})^{\frac{1}{2}}.\label{z2t}
		\end{empheq}
	\end{subequations}
	It is easy to show that 
	\begin{equation}\label{sumSquare1}
	z_{1}^{2} + z_{2}^{2}=1,
	\end{equation}
	which allows us to set
	\begin{subequations}\label{z1sinz2cos}
		\begin{empheq}[left=\empheqlbrace]{align}
			z_{1}&=\sin\phi(t);\label{z1tsin}\\	
			z_{2}&=\cos\phi(t).\label{z2tcos}
		\end{empheq}
	\end{subequations}
	\subsection{Modified Lotka-Volterra equations with the time-dependent parameters.}	
		
	\subsubsection{Time-dependent parameters}\label{for parameters are all time dependent}
	In the case that $\alpha$, $\beta$, $\gamma$, and $\delta$ are all constant, T-Function$^{\cite{T-Function}}$ may not be a solution for either $x_{1}$ or $x_{2}$ in Eq.($\ref{traditional L-V}$). This conundrum may be overcome by assuming that some of the four parameters are time dependent. We now discuss the behaviors of equations that involve the time-dependent parameters. First we modify Eq.($\ref{traditional L-V}$) as follows: 
	\begin{subequations}\label{L-V for all time dependent}
		\begin{empheq}[left=\empheqlbrace]{align}
		\frac{dx_{1}}{dt}&=\alpha(t)x_{1}(t) - \beta(t) x_{1}(t)x_{2}(t);\label{diff for prey all time dependent}\\	
		\frac{dx_{2}}{dt}&=-\gamma(t) x_{2}(t) + \delta(t) x_{1}(t)x_{2}(t),\label{diff for predator all time dependent}
		\end{empheq}
	\end{subequations}
	where $x_{1}$ referes to the preys and $x_{2}$ refers to the predators.
	Eq.($\ref{diff_form}$) also holds even if the parameters, $\alpha, \beta,\gamma, \delta,$ are time dependent. But this time, 
	\begin{center}
		\begin{equation}
		\int\frac{\gamma}{x_{1}}dx_{1}+\int\frac{\alpha}{x_{2}}dx_{2}-\delta x_{1}-\beta x_{2}=-\Lambda=\mathbf{constant}.
		\end{equation}
	\end{center}
	The carrying capacity is also defined as \(K=e^{-\Lambda}\), but this time it is equal to 
	\begin{equation}\label{carrying capacity time dependent parameters}
		K=e^{-\Lambda}=e^{\int\frac{\gamma}{x_{1}}dx_{1}}e^{\int\frac{\alpha}{x_{2}}d{x_{2}}}e^{-\delta x_{1}}e^{-\beta x_{2}}
	\end{equation}
	Similary to Eq.($\ref{coord trans}$), we may also introduce new coordinates $z_{1}(t)$ and $z_{2}(t)$ such that 
	\begin{subequations}\label{coord trans time}
		\begin{empheq}[left=\empheqlbrace]{align}
		z_{1}(t)&=\frac{1}{\sqrt{\Lambda}}(\delta x_{1} + \beta x_{2})^{\frac{1}{2}};\label{z1t time}\\	
		z_{2}(t)&=\frac{i}{\sqrt{\Lambda}}\biggl(\int\frac{\gamma}{x_{1}}dx_{1}+\int\frac{\alpha}{x_{2}}dx_{2}\biggr)^{\frac{1}{2}}.\label{z2t time}
		\end{empheq}
	\end{subequations}
	It is easy to demonstrate, as in Eq.($\ref{sumSquare1}$), that 
	\begin{equation}\label{sumSquare1 time}
	z_{1}(t)^{2} + z_{2}(t)^{2}=1,
	\end{equation}
	which means that even if the parameters are time dependent, we are also allowed to set, as in Eq.($\ref{z1sinz2cos}$),
	\begin{subequations}\label{z1sinz2cos time}
		\begin{empheq}[left=\empheqlbrace]{align}
		z_{1}(t)&=\sin\phi(t);\label{z1tsin time}\\	
		z_{2}(t)&=\cos\phi(t).\label{z2tcos time}
		\end{empheq}
	\end{subequations}
	Keeping in mind that all parameters and variables are time dependent, we drop $t$ in the following derivations for brevity. Further, because 
	\[
	\left\{	
	\begin{array}{ll}
	\int\frac{\gamma}{x_{1}}d{x_{1}}=\int\frac{\gamma}{x_{1}}\frac{dx_{1}}{dt}dt=\int\bigl[\frac{\gamma}{x_{1}}\bigl(\alpha x_{1}-\beta x_{1}x_{2}\bigr)\bigr]dt,\hspace{2pt}\mathrm{and}\\
	\int\frac{\alpha}{x_{2}}d{x_{2}}=\int\frac{\alpha}{x_{2}}\frac{dx_{2}}{dt}dt=\int\bigl[\frac{\alpha}{x_{2}}\bigl(-\gamma x_{2}+\delta x_{1}x_{2}\bigr)\bigr]dt,\hspace{2pt}\mathrm{as\hspace{2pt}well\hspace{2pt}as}
	\end{array}
	\right. 
	\]
	
	\[
	\left\{	
	\begin{array}{ll}
	\frac{d}{dt}\bigl(\int\frac{\gamma}{x_{1}}\frac{dx_{1}}{dt}dt\bigr)=\frac{\gamma}{x_{1}}\frac{dx_{1}}{dt}=\frac{\gamma}{x_{1}}\bigl(\alpha x_{1}-\beta x_{1}x_{2}\bigr)\\
	\frac{d}{dt}\bigl(\int\frac{\alpha}{x_{2}}\frac{dx_{2}}{dt}dt\bigr)=\frac{\alpha}{x_{2}}\frac{dx_{2}}{dt}=\frac{\alpha}{x_{2}}\bigl(-\gamma x_{2}+\delta x_{1}x_{2}\bigr),
	\end{array}
	\right. 
	\]
	we get
	\[
	\dot{z_{2}}=\frac{i}{\sqrt{\Lambda}}\frac{\frac{\gamma}{x_{1}}\frac{dx_{1}}{dt}+\frac{\alpha}{x_{2}}\frac{dx_{2}}{dt}}{2\bigl(\int\frac{\gamma}{x_{1}}dx_{1}+\int\frac{\alpha}{x_{2}}dx_{2}\bigr)^{1/2}}.
	\]
	Therefore, 
	\begin{equation}\label{gammaz122z2z2dot}
	\begin{aligned}
		\gamma z_{1}^{2}-2z_{2}\dot{z_{2}}
		&=\frac{\gamma}{\Lambda}(\delta x_{1}+\beta x_{2})
				-2\frac{i}{\sqrt{\Lambda}}\biggl(\int\frac{\gamma}{x_{1}}dx_{1}+\int\frac{\alpha}{x_{2}}dx_{2}\biggr)^{1/2}\frac{i}{\sqrt{\Lambda}}\frac{\frac{\gamma}{x_{1}}\frac{dx_{1}}{dt}+\frac{\alpha}{x_{2}}\frac{dx_{2}}{dt}}{2\bigl(\int\frac{\gamma}{x_{1}}dx_{1}+\int\frac{\alpha}{x_{2}}dx_{2}\bigr)^{1/2}}\\
		&=\frac{\gamma}{\Lambda}(\delta x_{1}+\beta x_{2})+\frac{1}{\Lambda}\bigl[\frac{\gamma}{x_{1}}\bigl(\alpha x_{1}-\beta x_{1}x_{2}\bigr)+\frac{\alpha}{x_{2}}\bigl(-\gamma x_{2}+\delta x_{1}x_{2}\bigr)\bigr]\\
		&=\frac{1}{\Lambda}\bigl(\gamma\delta x_{1}+\gamma\beta x_{2}+\gamma\alpha-\gamma\beta x_{2}-\alpha\gamma+\alpha\delta x_{1}\bigr)\\
		&=\frac{1}{\Lambda}\delta x_{1}(\alpha+\gamma).
	\end{aligned}
 	\end{equation}
 	Similarly, we may also obtain
 	\begin{equation}\label{alphaz122z2z2dot}
 		\alpha z_{1}^{2}+2z_{2}\dot{z_{2}}=\frac{1}{\Lambda}\beta x_{2}(\alpha+\gamma).	
	\end{equation}
	Imitating Evans and Findley's idea$^{\cite{Evans and Findley}}$, we may also define 
	\begin{equation}\label{omega time}
		\omega(t)=\Lambda\cdot\frac{\sin^{2}\phi(t)}{\alpha(t)+\gamma(t)}.
	\end{equation}
	Therefore, 
	\begin{equation*}
	\begin{aligned}
		\dot{\omega}(t)
		&=\frac{\Lambda}{(\alpha+\gamma)^{2}}\bigl[2\sin\phi\cos\phi\dot{\phi}(\alpha+\gamma)-(\dot{\alpha}+\dot{\gamma})\sin^{2}\phi\bigr]\\
		&=\frac{\Lambda}{\alpha+\gamma}\sin(2\phi)\dot{\phi}-\frac{\Lambda}{(\alpha+\gamma)^{2}}(\dot{\alpha}+\dot{\gamma})\sin^{2}\phi.
	\end{aligned}
	\end{equation*}
	Therefore, 
	\begin{equation*}
	\begin{aligned}
		\gamma\omega+\dot{\omega}
		&=\gamma\Lambda\frac{\sin^{2}\phi}{\alpha+\gamma}+\frac{\Lambda}{\alpha+\gamma}\sin(2\phi)\dot{\phi}-\frac{\Lambda}{(\alpha+\gamma)^{2}}(\dot{\alpha}+\dot{\gamma})\sin^{2}\phi\\
		&=\frac{\Lambda}{\alpha+\gamma}(\gamma\sin^{2}\phi+\dot{\phi}\sin(2\phi))-\frac{\Lambda}{(\alpha+\gamma)^{2}}(\dot{\alpha}+\dot{\gamma})\sin^{2}\phi.
	\end{aligned}
	\end{equation*}
	But \( \gamma z_{1}^{2}-2z_{2}\dot{z_{2}}=\gamma\sin^{2}\phi+\dot{\phi}\sin(2\phi)\), after combining Eq.($\ref{gammaz122z2z2dot}$) and Eq.($\ref{omega time}$), we obtain
	\begin{equation}
		\gamma\omega+\dot{\omega}=\delta x_{1}-\frac{\dot{\alpha}+\dot{\gamma}}{\alpha+\gamma}\omega.
	\end{equation}
	Similary derivations may lead to
	\begin{equation}
		\alpha\omega-\dot{\omega}=\beta x_{2}+\frac{\dot{\alpha}+\dot{\gamma}}{\alpha+\gamma}\omega.
	\end{equation}
	Thus,
	\begin{subequations}
		\begin{empheq}[left=\empheqlbrace]{align}
		x_{1}&=\frac{1}{\delta}\bigl[\omega\bigl(\gamma+\frac{\dot{\alpha}+\dot{\gamma}}{\alpha+\gamma}\bigr)+\dot{\omega}\bigr]=\frac{1}{\delta}\bigl(\omega\Gamma+\dot{\omega}\bigr);\\
		x_{2}&=\frac{1}{\beta}\bigl[\omega\bigl(\alpha-\frac{\dot{\alpha}+\dot{\gamma}}{\alpha+\gamma}\bigr)-\dot{\omega}\bigr]=\frac{1}{\beta}\bigl(\omega A-\dot{\omega}\bigr),
		\end{empheq}
	\end{subequations}
	where
	\begin{subequations}
		\begin{empheq}[left=\empheqlbrace]{align}
		\Gamma&=\gamma+\frac{\dot{\alpha}+\dot{\gamma}}{\alpha+\gamma};\\
		     A&=\alpha-\frac{\dot{\alpha}+\dot{\gamma}}{\alpha+\gamma},
		\end{empheq}
	\end{subequations}
	from which we may deduce
	\begin{equation}
		\delta x_{1}+\beta x_{2}=\omega(\alpha+\gamma).
	\end{equation}
	If the four time-dependent parameters are not varrying dramatically with time, we may still make use of Eq.($\ref{carrying capacity}$) as the carrying capacity and Eq.($\ref{maximum carrying capacity}$) as the maximum carrying capacity. It should not be surprising that under this assumption, some of the parameters may not be positive. On the other hand, it is mearly a pompous thought to consider that human race can only play the role as the predator. Completely opposite to this, before the modern time, we humans had been serving as the preys of some gigantic predators for a very long time, such as lions, tigers, leopards, and crocodilians. Bears, Komodo dragons and hyenas may also eat humans whenever they had chances.$^{\cite{man-eater}}$. Even for the present time, diseases, wars, and polutions may also be regarded as the predators that may do harm to human beings.\\
	Keeping this in mind, we believe that the best way to make use of the Lotka-Volterra equations could be that we divide the situations into two scenarios. The first one is that we treat human population as the prey and all the lethal factors that may jeopardize the existence of humans as the predator. On the contrary, in the second scenario, we adopt the humans as the predators and the natural resources as the preys.
	
	\subsubsection{Scenario 1: Human population vs. lethal factors as the prey and the predator}\label{Case 1} 
\begin{center}
		\begin{figure}[htbp]
			\centering
			\includegraphics[width=0.8\textwidth]{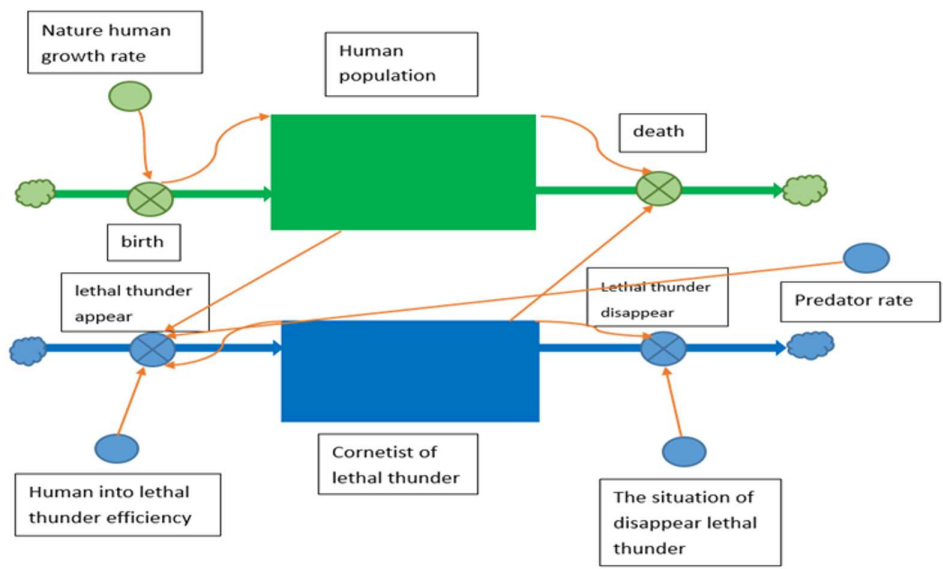}
			\caption{The relationship between lethal factors (predators) and humans (preys).}
			\label{fig:relationshipLethalFactorsHumans}
		\end{figure}
	\end{center}

	In this case, we treat that $x_{1}$ is the human population and $x_{2}$ is the lethal factor at time $t$, respectively, whose interactive relationship may be illustrated in Figure \ref{fig:relationshipLethalFactorsHumans}.
	Instead of assuming that all parameters are constants as in Sec.(\ref{L-V Section}), we ease the restrictions on allowing $\gamma$ and $\delta$ to be time dependent with the relation $\delta(t) = \zeta\gamma(t)$, where $\zeta$ is a constant, while still maintaining $\alpha$ and $\beta$ to be constant. By doing so, we acquire the modified Lotka-Volterra equations as follows:
	\begin{subequations}\label{L-V for Case 1}
		\begin{empheq}[left=\empheqlbrace]{align}
		\frac{dx_{1}}{dt}&=\alpha x_{1}(t) - \beta x_{1}(t)x_{2}(t);\label{diff for prey Case 1}\\	
		\frac{dx_{2}}{dt}&=-\gamma(t) x_{2}(t) + \delta(t) x_{1}(t)x_{2}(t)
		                 =-\gamma(t) x_{2}(t) + \zeta\gamma(t) x_{1}(t)x_{2}(t)\label{diff for predator Case 1}
		\end{empheq}
	\end{subequations}

	Making use of the fact that the T-Function fits the recorded human population quite well$^{\cite{T-Function}}$, it is natural to think that the prey $x_{1}$ satisfies 
	\begin{equation}
	x_{1}(t)=\frac{A}{\lbrack\ln{(B+E)}\rbrack^{M}},\label{T-Function x1}
	\end{equation}
	where $E$ is a function of t such that \(E(t)=e^{\frac{D-t}{\tau}}\). The parameters in Eq.(\ref{T-Function x1}) were already given in Ref\cite{T-Function} with $A=3.83$ billion, $B=1.28$, $D=1980$, $M=0.70$, and finally $\tau=22.9$ years. We now study the behavior of $\gamma(t)$. First, \[\frac{dx_{1}}{dt}=x_{1}\cdot\frac{ME}{\tau (B+E)\ln{(B+E)}}=x_{1}\cdot(\alpha-\beta x_{2}).\] Therefore, 
	\begin{equation}\label{x2 in Case 1}
	x_{2}(t)=\frac{\alpha}{\beta}-\frac{ME}{\beta\tau(B+E)}\ln{(B+E)}=\eta-\frac{ME}{\beta\tau(B+E)\ln{(B+E)}},
	\end{equation}
	assuming \(\frac{\alpha}{\beta}=\eta.\) Simple derivations of differentiating the above equation with respect to $t$ gives us 
	\[\frac{dx_{2}}{dt}=\frac{ME\lbrack B\ln(B+E)-E\rbrack}{\beta\tau^{2}(B+E)^{2}\lbrack\ln(B+E)\rbrack^{2}}.\] Combining the above equation with Eq.(\ref{diff for predator Case 1}), we may obtain 
	\begin{equation}\label{gamma(t)}
	\gamma(t)=\frac{ME\lbrack \ln(B+E)\rbrack^{M-1}\lbrack B\ln(B+E)-E\rbrack}{\tau (B+E)\lbrace \zeta A\lbrack\eta\tau\beta(B+E)\ln{(B+E)}-ME\rbrack + ME\lbrack\ln{(B+E)}\rbrack^{M} -\eta\tau\beta(B+E)\lbrack\ln{(B+E)}\rbrack^{M+1}\rbrace}.
	\end{equation}

	\subsubsection{Scenario 2: Natural resources vs. Human Population as the prey and the predator}\label{Case 2} Next, we consider an opposite situation at which $x_{2}$ is assumed to be the human population (the predator), also described by the T-Function, while $x_{1}$ the natural resources (the prey). Under this scenario, we assume that $\beta(t)=\zeta\alpha(t)$ while $\gamma$ and $\delta$ are still constants, comparing against assuming $\alpha$, $\beta$ as constants and $\gamma$, $\delta$ as time-varying functions in Sec.(\ref{Case 1}). Thus, we acquire another new form of the modified Lotka-Volterra equations as follows:
	\begin{subequations}\label{L-V for Case 2}
		\begin{empheq}[left=\empheqlbrace]{align}
		\frac{dx_{1}}{dt}&=\alpha(t) x_{1}(t) - \beta(t) x_{1}(t)x_{2}(t)=\alpha(t) x_{1}(t) - \zeta\alpha(t) x_{1}(t)x_{2}(t);\label{diff for prey Case 2}\\	
		\frac{dx_{2}}{dt}&=-\gamma x_{2}(t) + \delta x_{1}(t)x_{2}(t).\label{diff for predator Case 2}
		\end{empheq}
	\end{subequations}
	Setting 	
	\begin{equation}
	x_{2}(t)=\frac{A}{\lbrack\ln{(B+E)}\rbrack^{M}},\label{T-Function x2}
	\end{equation} with similar derivations as in Sec.(\ref{Case 1}), we may obtain the following formula.
	First, 
	\[\frac{dx_{2}}{dt}=x_{2}\cdot\frac{ME}{\tau (B+E)\ln{(B+E)}}=x_{2}\cdot(-\gamma+\delta x_{1}).\] Therefore, 
	\begin{equation}\label{x1 in Case 2}
	x_{1}(t)=\frac{\gamma}{\delta}+\frac{ME}{\delta\tau(B+E)}\ln{(B+E)}=\eta+\frac{ME}{\delta\tau(B+E)\ln{(B+E)}},
	\end{equation}
	assuming \(\frac{\gamma}{\delta}=\eta.\) Likewise, simple derivations of differentiating the above equation with respect to $t$ gives us 
	\[\frac{dx_{1}}{dt}=\frac{ME\lbrack E-B\ln(B+E)\rbrack}{\delta\tau^{2}(B+E)^{2}\lbrack\ln(B+E)\rbrack^{2}}.\] Also, combining Eq.(\ref{diff for prey Case 2}) gives us
	\begin{equation}\label{alpha(t)}
	\alpha(t)=\frac{ME\lbrack \ln(B+E)\rbrack^{M-1}\lbrack E-B\ln(B+E)\rbrack}{\tau (B+E)\lbrace \zeta A\lbrack\eta\tau\delta(B+E)\ln{(B+E)}+ME\rbrack - ME\lbrack\ln{(B+E)}\rbrack^{M} -\eta\tau\delta(B+E)\lbrack\ln{(B+E)}\rbrack^{M+1}\rbrace}.
	\end{equation}
	\section{Results and Discussions}
	In this section, we first discussed factors that may have influence on the carrying capacity on Earth. Then we summarized the parameters that are used in either Scenario 1 or Scenario 2. Starting from this we showed the plots of T-Function, together with the fitted observed data, the carrying capacity, as well as some time-varying functions.
	\subsection{The major factors that restrict the carrying capacity of Earth}
\begin{center}
		\begin{figure}[htbp]
			\centering
			\includegraphics[width=0.7\textwidth]{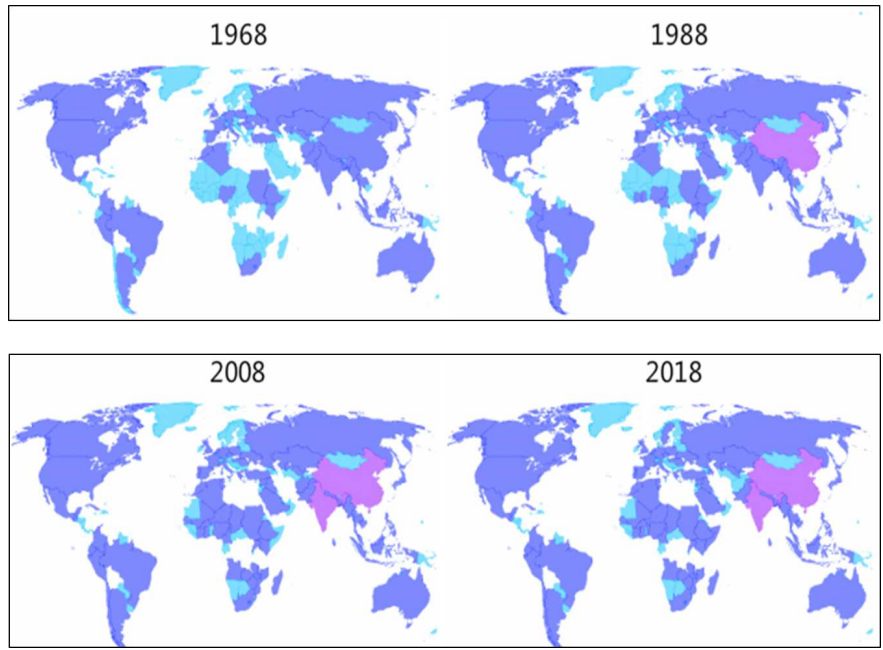}
			\caption{The world population map from 1968 to 2018.}
			\label{fig:world Pop Map}
		\end{figure}
	\end{center}

With the behavior of humankind, we are facing the problem of the carrying capacity reaching the peak soon. There were many issues with nature nowadays such as water shortage, the decline of land fertility, reduction of the forest area, and the diminishing of the cultivated land.  We understand that humankind only concerns about the development of technology but ignores the side effects of what we have done previously. The application of water resources is manifold. They are compulsory to the survival of human species as well as the development of agriculture. Nevertheless, freshwater resources are limited on Earth. Though Earth is regarded as the "blue planet" for it is covered with water practically, more than 95\% of them are not able to be utilized, plus it is troublesome to achieve freshwater from the sea. Relatively, humans can barely survive on less than 5\% of the water resources now. Per capita of water resources would decline for the population explosion. We must face the obstacle of the lack of water resources in the prospect if freshwater has achieved a certain proportion.\\ 
	What is more, oxygen is another mandatary factor for mankind to survive on Earth. The respiration of both humans and trees affects the proportion of the balance of carbon dioxide and oxygen. Recently, the number of trees has shown a declining trend for the development of mankind. Wood resources are essential to the construction of architecture. Hence, humanity has cut more trees to finish it casually. Plus, trees were cut if humankind wanted to expand their living quarters. As deforestation happened, the absorption of carbon dioxide would decrease as well. The balance of carbon dioxide and oxygen in the air would be devastated if the population keeps increasing while number the trees is reducing. Regarding the message, the banishment of the forest area is severe for six football field of forest vanishing within a minute. Consequently, both water and oxygen resources have severe impacts on Earth's carrying capacity.\\
	Apart from the necessity of freshwater, arable land is an essential part of the development of agriculture. Food is fundamental to the survival of humankind. Humans must sustain their life; therefore, they have done the reclamation unrestrained aimed to survive. Immensely using the cultivated land, it cannot deal with the supply of fertility. Under natural conditions, arable land must stop farming after farming plants for a while. Humanity skipped the process for they want to gain more food resources. Two primary purposes are survival and gaining benefits from food resources. The cultivated field has overworked and offered fertility all the time. Consequently, fertility in the soil will decline in a normal situation. Moreover, using chemical fertilizers against natural law is a way to boost profits. However, the soil is polluted at the same time. Humankind gained much food by using chemical fertilizers, and the pollution of the land has become severe. In addition to fertility decline, acidification of the soil is dangerous. It devastated the land structure, and it brought quite a bit of harm to the living circumstance of mankind. In the process of human reproduction, fertilizers produced large amounts of organic acids that cause acidification to the land. Plus, whether the cultivated land can be used is relevant to human survival. If the population grows, the per capita footprint gradually decreases. More and more construction for residence is needed. Besides, the advantages of real estate are much more than agriculture. Humankind would instead use cultivated land as a place of construction because of profit relations. There are fewer and fewer acres of cultivated land on the earth because of the activity of mankind for we have devastated the soil and generate an irreversible impact. We used cultivated land selfishly, and this would make the earth's carrying capacity reaching an end.\\ 
	The ocean, although it is salt water, is vital to our human life. On our "blue home," we understood that the area of the sea is larger than that of land. Likewise, the number of species living in the sea is more than the number of species living on the ground. According to the statistic, the earth had an extinction crisis before 250 million years, and all creatures living on Earth had extinct at one night. Meanwhile, scientists consider that the ocean acidification caused by rising carbon dioxide in the air may need to be blamed for the extinction of mankind in the future. It was worth noting that the industrial revolution made the PH value of seawater dropping because of carbon dioxide.\\
	Furthermore, carbon dioxide increases because humans burn fossil fuels in large quantities. In the opinion of the predecessor, the draft will absorb all the carbon dioxide after the emission. The ocean has a self-sustaining method of acid-base balance. Carbonic acid in terrestrial rocks and seabed rocks and the carbon dioxide formed by the dissolution of atmospheric carbon dioxide in seawater will occur a reaction together. They are thereby achieving the role of acid-base balance. About 40\% of the emission of carbon dioxide was being absorbed by the ocean. It requires to spend hundreds of thousands of years to purify carbon dioxide that has been absorbed by seawater. The severe acidification of the ocean has caused much death of the lives in the sea, such as fish, coral, and shellfish. Moreover, they are the other main food for human. Humankind would suffer from a shortage of food if ocean acidification continued to intensify. Whether it is water resources, forest resources, or the cultivated land are all important. We could assume their existence equal to the presence of human life. Conversely, the human beings will die if they vanished. 
	Uneven distribution of the world's population was an enormous issue as shown in Figure \ref{fig:world Pop Map}$^{\cite{kaggle}}$; the issue occurred from ancient to now. For if a region had an overpopulation issue, it led to roar property prices and decline the living standard. As a result, it would influence the population in that region. Therefore, the uneven distribution of the world's population was also the one of the main reasons for the carrying capacity of the earth accelerates to the limit due to the problem of arable land, acidification of the ocean, water shortage or reduction of the forest area.
	
	\subsection{Parameters and the calculated carrying capacity}
\begin{center}
		\begin{figure}[htbp]
			\centering
			\includegraphics[width=0.8\textwidth]{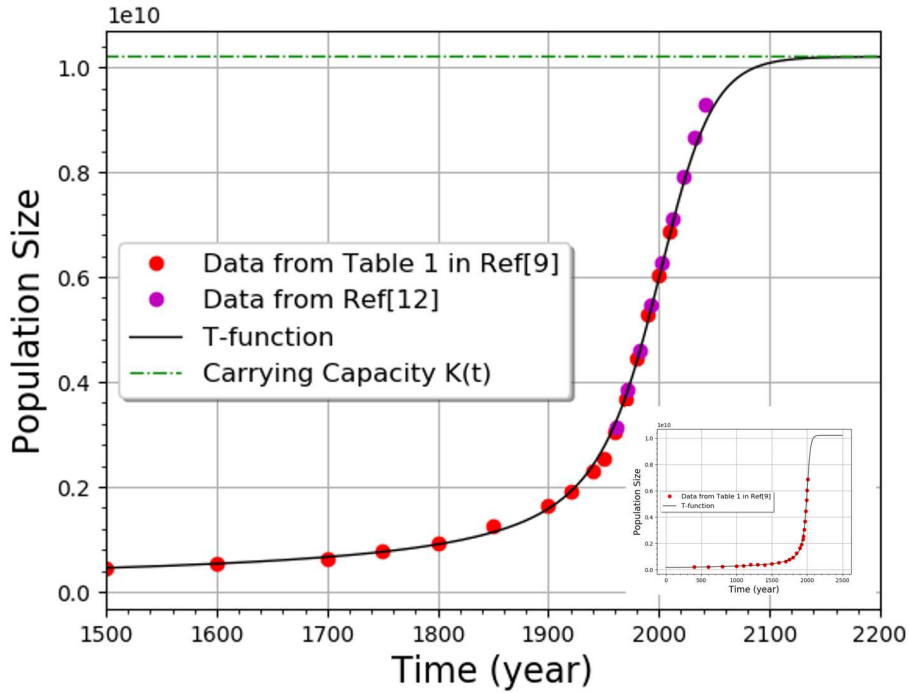}
			\caption{T-function in shorter duration of time, comparing with fitted data from two sources. $K(t)$ refers to the carrying capacity, which is a constant function in time with value $10.2$ billion. (Inlet) Plotting of the whole-range T-Function ranging from Year 0 to Year 2200.}
			\label{fig:allTogetherAndTfunction}
		\end{figure}
	\end{center}

	Table \ref{tab:alphatozeta}	summarized the parameters we used in Sec.(\ref{Case 1}) and in Sec.(\ref{Case 2}).  It should be clear that in Scenario 1, $\alpha$ and $\beta$ are constants while $\gamma$ and $\delta$ are time-dependent; whereas in Scenario 2, $\alpha$ and $\beta$ are time-dependent while $\gamma$ and $\delta$ are constants. $\eta$ and $\zeta$ are the same in both Scenarios. Moreover, $\alpha$ and $\beta$ in Scenario 1 correspond to $\gamma$ and $\delta$ in Scenario 2, respectively.\\
	Figure \ref{fig:allTogetherAndTfunction} shows the T-Function in Eq.(\ref{T-Function x1}), together with two recorded data in references$^{\cite{T-Function},\cite{kaggle}}$, and the calculated Carrying Capacity in Eq.(\ref{carrying capacity}). The main idea of choosing parameters was to match the value of the maximum carrying capacity with that of the saturated T-Function.\\
	As we observe, the carrying capacity is time independent, with the constant value roughly equal to 10.2 billion. It seemed to make sense because the carrying capacity is a fixed number once the surrounding conditions in the environment (comparable to the parameters in Table \ref{tab:alphatozeta} in our study) are all set. In other words, the carrying capacity of human population has been determined at the very beginning of the birth of Earth.
	
	\begin{center}
		\begin{table}[htbp]
			\centering
			\begin{tabular}{*{3}{c}} 
			\hline\hline
	Parameters & Scenario 1 in Sec.($\ref{Case 1}$)   & Scenario 2 in Sec.($\ref{Case 2}$)             \\
			\hline
    $\alpha$&      23.0457                          &    Eq.($\ref{alpha(t)}$)                     \\
	$\beta $&  $\frac{\alpha}{e^{2}}\approx3.1189$  &  $\beta(t)=\zeta\alpha(t)$                   \\
	$\gamma$&  Eq.($\ref{gamma(t)}$)                &  23.0457                                     \\
	$\delta$& $\delta(t)=\zeta\gamma(t)$            &  $\frac{\gamma}{e^{2}}\approx3.1189$         \\
	$\eta $ & $\frac{\alpha}{\beta}=e^{2}\approx7.3891$ &$\frac{\gamma}{\delta}=e^{2}\approx7.3891$\\
	$\zeta$ &  25                                   &  25                                          \\
			\hline\hline	
			\end{tabular}
			\caption{Parameters that are used in either Scenario 1 in Sec.($\ref{Case 1}$) or Scenario 2 in Sec.($\ref{Case 2}$).}
			\label{tab:alphatozeta}
		\end{table}
	\end{center}

	\subsection{The Lethal factors and $\gamma(t)$ as the time-varing functions}

\begin{center}
		\begin{figure}[htbp]
			\centering
			\includegraphics[width=0.8\textwidth]{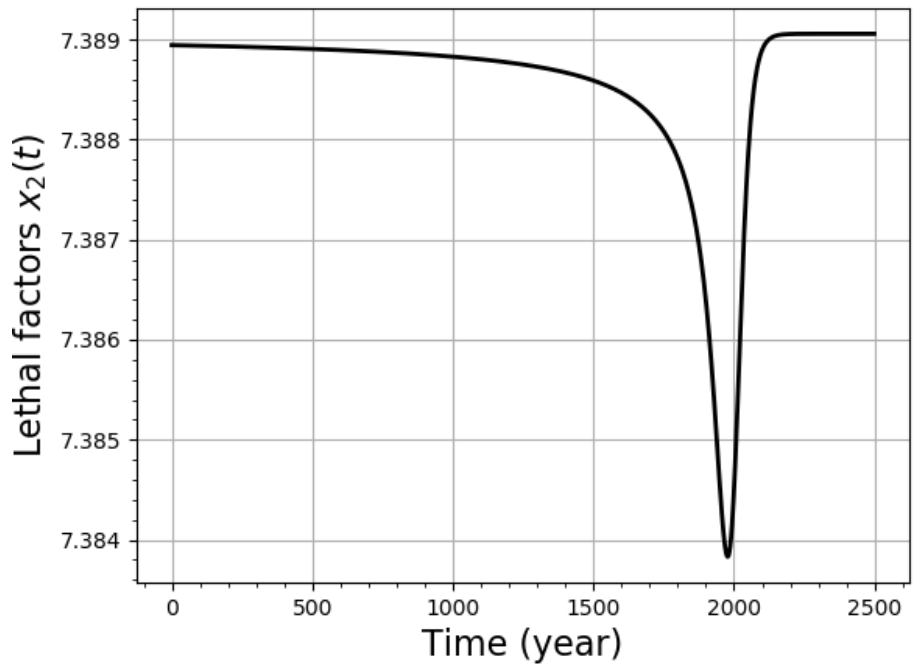}
			\caption{Plotting of $x_{2}(t)$.}
			\label{fig:lethal factors}
		\end{figure}
	\end{center}

\begin{center}
		\begin{figure}[htbp]
			\centering
			\includegraphics[width=0.8\textwidth]{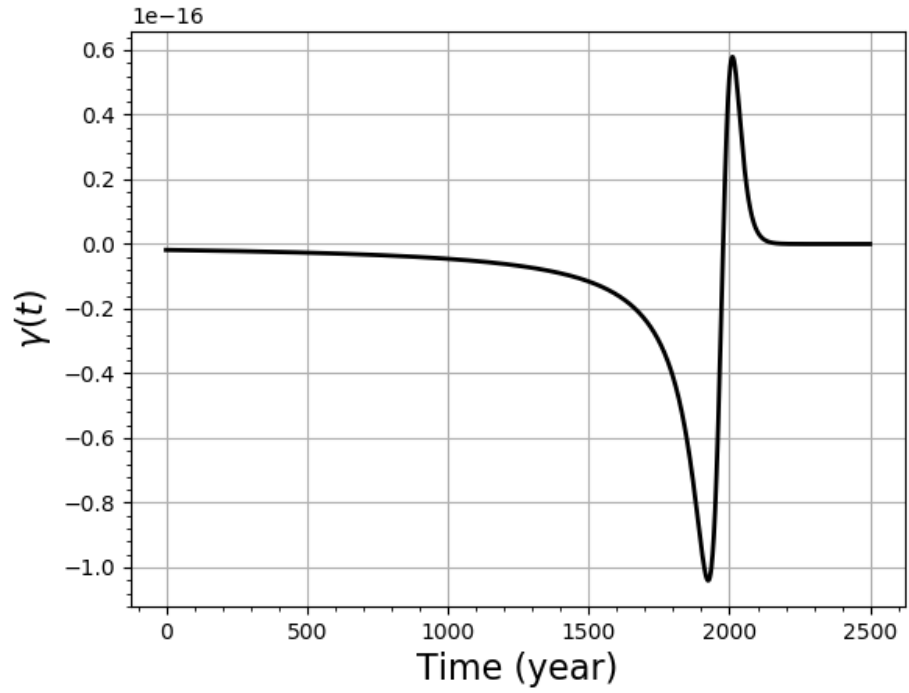}
			\caption{Plotting of Eq.(\ref{gamma(t)}) for $\gamma(t)$.}
			\label{fig:gamma}
		\end{figure}
	\end{center}

	In Figure \ref{fig:lethal factors}, we understand that the predator is the value of the lethal factors that would lead to negative effects on the human population. From Year 0 to approximately Year 2000, we notice that the predator had been significantly decreasing down to the value lower than 7.384. Nevertheless, around Year 2000, the predator had been sharply increasing and reaching at 7.389 about Year 2100, remaining stable at the value even upto Year 2500. The trend of this lethal function with time may be explained by stating that before Year 2000, attributed to the development of technology and the improvement of living circumstances, we humans had been reducing the "lethal factors". However, after Year 2000, even if our technology and society still keep improving and developing, the lethal factors have been increase rather than decreasing, possibly due to the misuse or overuse of technology and our natural resources, raising the amount of the undesirable and unpleasant harmful substances on Earth. 
	As shown in Figure \ref{fig:gamma}, $\gamma(t)$ is a function with quite small numbers, roughly ranging from $-1.0\times10^{-16}$ to $0.6\times10^{-16}$. The overall description about the shape of the function may be that it decreased from Year 0 to Year 1900 down to the value $-1.0\times10^{-16}$, and changed steeply (with only small amount) around that year, climbing up to the value $0.6\times10^{-16}$ around Year 2000, and then returned and remained to roughly zero. 
	
	\subsection{The natural resources and $\alpha(t)$ as the time-varing functions}
\begin{center}
		\begin{figure}[htbp]
			\centering
			\includegraphics[width=0.8\textwidth]{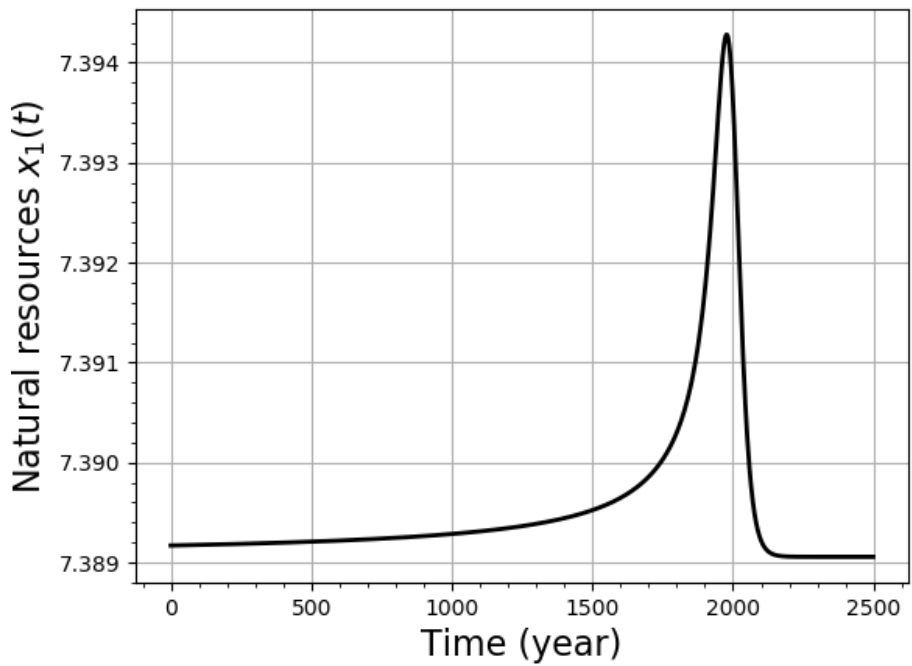}
			\caption{Plotting of $x_{1}(t)$.}
			\label{fig:natural resources}
		\end{figure}
	\end{center}

From Figure \ref{fig:natural resources}, we discovered that the natural resources had been gradually increasing from Year 0 to Year 1500. This could mean that we had been excavating and making use of more and more natural resources during the period. Afterwards, it led to a significantly abrupt increasing between Year 1500 and Year 1950. Unfortunately, it had a dramatic drop from Year 1950 to Year 2100. We may explain this by observing the fact that we almost run out of the natural resources in modern times. Then, the natural resource will be staying stable thereafter.
	
\begin{center}
		\begin{figure}[htbp]
			\centering
			\includegraphics[width=0.8\textwidth]{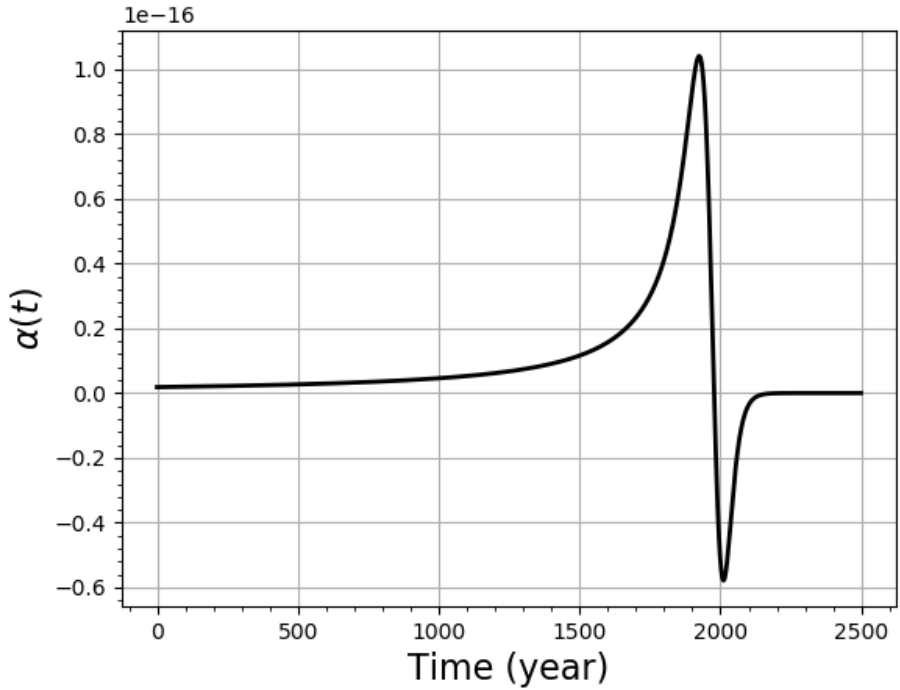}
			\caption{Plotting of Eq.(\ref{alpha(t)}) for $\alpha(t)$.}
			\label{fig:alpha}
		\end{figure}
	\end{center}

	It is evident that the impact we discovered here was the same as the one we found in the previous subsection. We perceived that abrupt changes happened around 2000.
	For our perspective, we considered that it may be because during the course of time, the natural resources have be exploited, diminishing the lethat factors as well, as shown in Figure \ref{fig:lethal factors}.
	Because we assumed the population as the predator. Therefore, we needed to acknowledge that the result of the human population in this sub-section should be the same as the result of the population in the previous sub-section. From the inlet in Figure \ref{fig:allTogetherAndTfunction}, we knew that the human population had increased to approximately 11 billion between Year 2000 and Year 2500. Hence, we had the same result for the human population in our model, which meant that the result for the calculations on the natural resources with respect to time would be quite reasonable.
	As shown in Figure \ref{fig:alpha}, $\alpha(t)$ kept enhancing approximately from $0.04\times10^{-16}$ to $1.4\times10^{-16}$ between Year 0 and Year 1900. Unexpectedly, it dramatically declined from $1.4\times10^{-16}$ to $-0.56\times10^{-16}$ in between Year 1900 and Year 2000. Based on this unearthing of $\alpha(t)$, we discovered that the changing of $\alpha(t)$ is similar to $\gamma(t)$ (Figure \ref{fig:gamma}) and $x_{1}(t)$ (Figure \ref{fig:natural resources}). They all had a significant changing around Year 2000.
	It is obvious that Figure \ref{fig:lethal factors} and Figure 
	\ref{fig:natural resources} are symmetric to the axis $y=7.389$, whereas Figure \ref{fig:gamma} and Figure \ref{fig:alpha} are symmetric to the axis $y=0.0$. Also, comparing through Figure \ref{fig:lethal factors} to Figure \ref{fig:alpha}, it is evident that the dramatic changes occurred in almost same year around 1900 to 2000. It is interesting what might happen during 1900 to 2000 for the dramatic differences to occur. To our first thought, it is an astonishing coincidence with several religious prophecies that near Year 2000 some big events may have happened. For instance$^{\cite{near Y2K}}$, millennialists believe that in that period of time, subsequent with the final judgment and future eternal state of the "World to Come", the earth would welcome a Golden Age or Paradise. Similar to this, Zoroastrianism states that each continual 1000-year period of time ends in a total annihilation of destruction and cataclysm until the end of the final millennial age, which is belived to be Year 2000, arrived and a triumphant king of peace won the fight against the evil spirit. 
	Although most of the authors do not have religious belief, we still believe that Year 2000 is a turning point of human race. As a matter of fact, it is because of human activities that cause most of the global warming. There is higher and higher probability that we have extreme climates, all of which may be blamed for over-exploitation of natural resources by humans. Technology development could only deteriorate the situation rather than providing solutions to the problem.
	\subsection{Factors on limiting the carrying capacity}
	Nature should be the bigest limitation of the world’s carrying capacity because it is irresistible that humans are affected by nature itself in development and distribution. Thus, the natural factors are chosen as the crucial factors in our discussion. After having this concept, land resource and energy problem are the first elements to be considered.\\ 
	However, precautious considerations make us to choose the energy as the major factor in limiting the carrying capacity. The reason for us to leave behind the land resource is that it can be effectively solved by advance urban planning like the Singaporean Government does. The population of Singapore was 6,028,912$^{\cite{SingaporeCountryMeter}}$ at the present time and the area of Singapore is approximately 721.571 square kilometers$^{\cite{SingaporeWiki}}$. A simple circulation, we got the population density of Singapore is about 8355.258 per square kilometer. The world population is approximately 7,714,576,923$^{\cite{WorldPopulationProjections}}$ at the moment. If we use the urban planning of Singapore to accommodate the world population, we may get that we can only use about 923,320 square kilometers for all the people in the world and that is about 9.1\% of the area of Europe.\\
	On the other hand, energy is an essential requirement nowadays and even in the future because it may be runing out soon in the high demand situation right now. Moreover, the renewable and the new energy resource are still at the very beginning of the development and application, which means that we still have a long time to go in order to replace the limited energy like coal or oil with them.
	\begin{center}
		\begin{table}[htbp]
			\centering
			\begin{tabular}{*{4}{c}} 
				\hline\hline
				Year & Unrenewable (GWh) & Renewable(GWh) & Electricity Proportion\\
				\hline
				1990 &     9516471       &    2364747     &        13\%           \\
				1995 &	   10581821	     &    2727262	  &        14\%           \\ 
				2000 &     12550522	     &    2949714	  &        15\%           \\
				2005 &	  14935155	     &    3411933	  &        16\%           \\             
				2010 &	  17200288       &    4336559	  &        17\%           \\
				2015 &    18646323	     &    5689433	  &        18\%           \\
				2016 & 	  18925573	     &    6119497	  &        19\%           \\
				\hline\hline	
			\end{tabular}
			\caption{Data of the source used in electricity generation of the world in a different year from the International Energy Agency.$^{\cite{Electricity generation by fuel},\cite{Total Final Consumption}}$}
			\label{tab:energy}
		\end{table}
	\end{center}	
	Table \ref{tab:energy} illustrated the situation of the unrenewable source like coal gets a biger percentage comparing with the renewable source like hydro in electricity generation. Furthermore, the Electricity Proportion shows the percentage of electricity in total energy consumption which means that the renewable source is an uncommon use and plays a small role in the total expenditure.  Also, the efficiency of renewable energy is way less than the unrenewable energy, so the latter may still play an important role in the development of the world.
	\subsection{Possible ways for raising the carrying capacity of Earth}
	We consider developing underground urban that includes useful transportation of natural resources that may be the factor for raising the carrying capacity of the Earth.
	Based on the knowledge we have, we understand that carrying capacity is mainly calculated by available land area and total population, with the affection of natural resources, namely freshwater and agriculture, by which the output of this calculation will be affected. Hence, let us propose that if total population stays still and available land area is enlarged, upon which buildings are constructed, carrying capacity will possibly be increased. As the idea mentioned, we can achieve the ideal situation above by developing underground urban.\\
	As only 29.1 percent$^{\cite{infoplease}}$ of our planet is land, meaning that there is no enough space for 7.7 billion people to live. Ergo, developing underground city can expand our land area as a consequence of humans are allowed to build under the extraordinarily limited land. By detecting whether the underground contains any mineral products or any danger, namely magma, we are allowed to develop a city as it is safe to excavate a big area.\\  
	What is more, we also discover that the living place is also necessary for land use. Whence, urbanizing rural area can also reach the ideal situation we proposed. Of the greater importance, urbanization refers to a population shift from rural to urban where a significant amount of high-rising has built; this illustrates that living places for humans are developing to a higher position. By doing this, the problem of running out of living place will also be solved. It is not a rare phenomenon around the world. As claimed by some studies$^{\cite{emporis}}$, South Korea has the most significant amount of high-rising containing approximately 16,739 buildings in total. In addition, the country also possesses 229 skyscrapers. Let us take the tallest building in the world as another example: Burj Khalifa, located in Dubai has 163 floors$^{\cite{skyscrapercenter}}$. In this case, as we are allowed to build high-rising or skyscraper in urban, available living areas would increase while we keep building, the carrying capacity will also keep rising.\\
	On the other hand, Because of what we have done to our environment, the natural resources on Earth would be scarce in the near future that would severely affect the carrying capacity of the Earth. Hence, we need to discover more resources if we want to raise the carrying capacity. In this case, scientists have been exploring the space to find out the resources that we are lack of. In the same time, some scientists have found out plenty of metals from the Moon and other celestial bodies that we are short of. Moreover, they have also discovered that some asteroids have carbon-rich, metallic, or mineral-rich silicate. Since the number of minerals is diminishing nowadays, there will be fewer minerals that we can use for human life. Thus, scientists need to develop the space mining industry to discover more sources of minerals under the conditions of the development of technology.\\ 
	Furthermore, along with the amount of water with low salt content is diminishing nowadays, it means that there is less water that we can use to support human life in the future. Hence, there are being more wars for scrambling the water resources among the nations, take Kyrgyzstan and Uzbekistan for instance. With this in mind, if there are more countries that are short of the potable water, the carrying capacity of the Earth will reduce apparently because of the lack of resources. Also, the major support of human life is food. In this modern time, the relationship between population and food is not balanced. From 2014 to 2016, the number of people who are suffering from starvation is approximately 7.95 million. Meanwhile, the grain output in 2014 is 25.638 million ton, and there is 25.28 million ton in 2015. These two years are the most grain output of these years; however, there are still many people who are suffering from starvation under such much grain output, even in the future. It is evidence that the current grain output is not enough to support human’s demand. The main factor of it is the soil problem that is caused by the situation of over-cultivation, over-exploitation and the chemical fertilizer that make the fertility of the soil is decreasing. To sort out this problem fundamentally, people should have reasonable organic farming, take the traditional silkworm breeding method for example that could mix agriculture, fishery as well as the clothing industry together. Moreover, we need to promote diminishing the use of chemical fertilizers or using organic fertilizers 
	instead of chemical fertilizers.\\
	Another philosophical point of view is that we may inevitably wonder why we have to improve the carrying capacity for greater carrying capacity did not signify better living conditions on the earth. On the other hand, there are many other critical issues to deal with in the world, such as poverty, starvation, and low educational standard, etc. Those countries are mainly situated in Africa, particularly for the third-world countries; however, why do we need to raise the carrying capacity of the Earth, but not solve these problems at first? In support of this, there is a kind of rat that would commit collective suicide to produce more space for the whole family of the rats. If Earth had the issue of overpopulation, we may repeat the same mistake. We ought to learn nothing from past failures and make the same mistake. Consequently, we cannot help asking: is raising carrying capacity of the Earth the better idea to allow more humans to live in? 
	\subsection{Population hypothesis}
	In this subsection, we aimed to provide the hypothesis about humans living in the future after it had reached the peak of the population. Some authors$^{\cite{mice}}$ pointed out that having too many mice in one narrow living space would bring problems to the whole race. This gives us hints to getting to understand that the factors that would affect mice mortality are emigration, resource shortage, inclement weather, disease, and predation, as listed in Table \ref{tab:mice}.
	\begin{center}
		\begin{table}[htbp]
			\centering
			\begin{tabularx}{\textwidth}{c X}
				\hline\hline
				Factors & Explanations \\
				\hline
		Emigration &  In strange and less favorable habitats, the emigres become more exposed to other mortality factors.\\
		Resource shortage &	 Shortages of shelter, other environmental resources, and associates lead to debilitation and dissatisfaction of habitat that culminates in death\\ 
		Inclement Weather &  Any conditions of wind, rain, humidity or temperature which exceed the usual limits of tolerance which increases the risk of death through debilitation.\\
		Disease & Abnormally high densities enhance the likelihood of spread of disease to epidemic proportions.\\             
		Predation &	Through evolution, it would associate with its predators capable of killing some of its members.\\
		\hline\hline	
			\end{tabularx}
			\caption{Explanations of different factors which affect mice mortality.}
			\label{tab:mice}
		\end{table}
	\end{center}	
	Therefore, we originated in this success of introducing our hypothesis of the population. First and foremost, we discovered that the factors affecting the mice population are similar to those affecting the human population, raising the curiosity on what circumstainces we humans would face the same destiny as the mice. Humans would become rivals to each other, scrambling for the natural resources and the living materials, leading to the devastatingly dangerous situation for the whole human race. To sum up, in this subsection, we aimed to introduce the hypothesis of the human population in the future. We predicted that the population of human would decrease because of the factors in Table \ref{tab:mice}.
	\section{Conclusions}
	Of greatest importance, because of the rapid development of society, there are more and more problems that we have made to our living environment, namely for overpopulation, urbanization, global warming, as well as land desertification and other factors that would make our Earth to be overloaded. Because of these problems, more and more studies focus on discussing what  the major factors affect the carrying capacity of the Earth. Yet most of the researches did not provide the exact number of carrying capacity together with the direct factors because they did not have a specific mathematical model that can determine actual carrying capacity and factors.\\
	Unquestionably, with the understanding of the theories of T-function and modified Lotka-Volterra Equations, we successfully constructed the model to calculate the carrying capacity of Earth for humans. In our paper, we modified the Lotka-Volterra Equations with the assumption that two of the original four parameters in the traditional equations are time dependent. In the first place, we assumed that the human population (borrowed from the T-Function) plays the role as the prey while all lethal factors that jeopardize the existence of the human race as the predator. Although we could still calculate time-dependent lethal function, the idea of treating the prey as the lethal factors was too general, resulting in impratical to recognizing what lethal factors may be included. Hence, in the second part of the modified Lotka-Volterra Equations, we exchanged the roles between the prey and the predator. This time, we treated the prey as the natural resources while the predator as the human population (still borrowed from the T-Function). In this case, we successfully calculated the natural resources as a function of time, and we also determined that the carrying capacity of the Earth is 10.2 billion people in the current time.	Next, about the first part of the Lotka-Volterra Equations, we considered the interactions of humans and the limiting factors like prey and predator in the natural system to simulate the population growth affected by the other factors in order to illustrate a better estimation and find out the right parameters to calculate the $K$  value, comparing to the T-Function. The parameters $\alpha$, $\beta$, $\gamma$, and $\delta$ are to describe the interactions of two species, being identified that $\alpha$ and $\beta$ are related to the prey or human while $\gamma$ and $\delta$ are relate to the predator or bad factors. When we found the suitable value of each parameter, we would get a better cognition in the relationships between population growth and influences of factors.\\
	In addition, we used some methods to evaluate and analyze the factors that would lead the carrying capacity to get a decline. In the same way, we have also built three models with the three theories to ascertain the contemporary carrying capacity of Earth that is under the current conditions by measuring. Likewise, we have also determined what humans could do for increasing the Earth’s carrying capacity for human life. We discovered that different factors that would limit the Earth’s carrying capacity for humans based on the current conditions. However, after our analysis, we found that only energy resources would have the negative impact on the carrying capacity. We also discussed the factors that would raise the carrying capacity, and we discovered that urbanization which increased land use and the increase in living area could increase the carrying capacity.\\
	Last but not least, we identified that the solutions that we could take for arising the carrying capacity of our Earth in future conditions. As evidence, statistics shows that the relationships between the number of people who are suffering from starvation and the grain is getting imbalanced, and this might be an obstacle in the future. Therefore, with the conditions of the development of organic agriculture, people ought to attempt to reasonable cultivation as well as using the organic fertilizers comprehensively. In the same way, scientists should start to discover and explore some new resources to be utilized before the resources on Earth have run out of, like finding water on other planets and developing the space mining industry. After taking these measures, the Earth’s carrying capacity could someday have been promising raised under the conditions of significant development of organic cultivation together with technology for space exploration.
	
	\section{Acknowledgement}
	We thank Escola Choi Nong Chi Tai in Macao PRC for the kindness to support this research project. 
	
\end{document}